\documentclass[twocolumn,nofootinbib,superscriptaddress]{revtex4}%
\usepackage{color}
\usepackage{epsfig}
\usepackage{graphicx}
\usepackage{amsmath}
\usepackage{amsfonts}
\usepackage{amssymb}
\usepackage{epstopdf}
\usepackage{bbold}%
\usepackage{algorithm}
\usepackage{algorithmic}

\providecommand{\U}[1]{\protect\rule{.1in}{.1in}}
\begin{document}
\title{MCMC with Strings and Branes: The Suburban Algorithm}
\author{Jonathan J. Heckman}
\affiliation{Department of Physics, University of North Carolina at Chapel Hill, NC}
\author{Jeffrey G. Bernstein}
\affiliation{Analog Devices $\vert$ Lyric Labs, Cambridge, MA}
\author{Ben Vigoda}
\affiliation{Gamalon Labs, Cambridge, MA}

\begin{abstract}
Motivated by the physics of strings and branes, we introduce a general suite of Markov chain Monte Carlo (MCMC)
``suburban samplers'' (i.e., spread out Metropolis). The suburban algorithm involves an ensemble of statistical
agents connected together by a random network. Performance of the collective in reaching a fast and accurate inference depends primarily
on the average number of nearest neighbor connections. Increasing the average number of neighbors above zero initially leads to an increase in performance, though there is a critical connectivity with effective dimension $d_{\mathrm{eff}} \sim 1$,
above which ``groupthink'' takes over, and the performance of the sampler declines.
\end{abstract}
\maketitle

\vspace{-3mm}

\section{Introduction}

\vspace{-3mm}

Markov chain Monte Carlo (MCMC) methods are a remarkably robust way to sample
from complex probability distributions. Metropolis-Hastings (MH)
sampling \cite{Metropolis:1953am, Hastings:1970} stands out as an important
benchmark. An appealing feature of MH sampling is the
simple physical picture which underlies the general method. Roughly
speaking, the idea is that the thermal fluctuations of a particle moving in
an energy landscape provides a conceptually elegant way to sample from a target distribution.

But there are also potential drawbacks to MCMC methods.
For example, the  speed of convergence to the correct posterior is often unknown since a sampler can
become trapped in a metastable equilibrium for a long period of time. Once a
sampler becomes trapped, a large free energy barrier can obstruct an accurate determination of
the distribution. From this perspective it is
therefore natural to ask whether further inspiration
from physics can lead to new examples of samplers.

\begin{figure}[t!]
\centering
\includegraphics[
scale = 0.25, trim = 0mm 0mm 0mm 0mm
]
{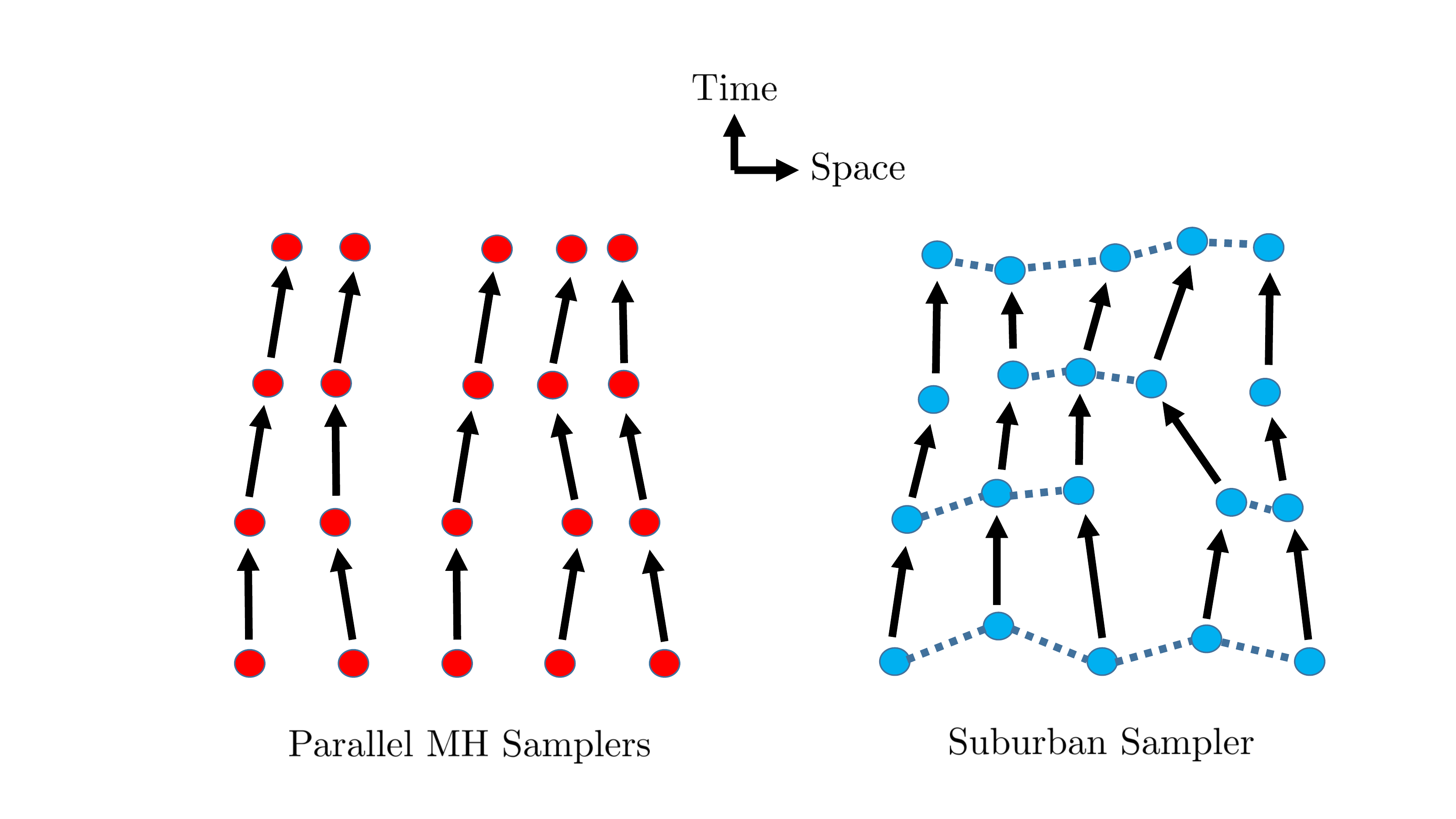}\caption{Depiction of how parallel MH\ samplers (left) and
a suburban sampler (right) evolve as a function of time.}%
\label{parstringbrane}%
\end{figure}

Now, although the physics of point particles underlies much of our modern
understanding of natural phenomena, it has proven fruitful
to consider objects such as strings and branes with finite extent in
$p$ spatial dimensions (a string being a case of a $1$-brane). One of the main features of branes is
that the number of spatial dimensions strongly affects how a localized perturbation
propagates across its worldvolume. Viewing a brane as a collective of point
particles that interact with one another (see fig. \ref{parstringbrane}),
this suggests applications to questions in statistical inference \cite{Heckman:2013kza}.

Motivated by these physical considerations, our aim in this work will be to
study generalizations of the MH algorithm for such extended
objects. For an ensemble of $M$ parallel MH samplers of a distribution
$\pi(x)$, we can alternatively view this as a single particle sampling from
$M$ variables $x_{1},...,x_{M}$ with density:
\begin{equation}
\pi(x_{1},...,x_{M})=\pi(x_{1})...\pi(x_{M}),
\end{equation}
where the proposal kernel is simply:
\begin{equation}
q_{\text{par}}(x_{1}^{\text{new}},...,x_{M}^{\text{new}}|x_{1}%
^{\text{old}},...,x_{M}^{\text{old}})= \underset{\sigma=1}{\overset{M}{%
{\displaystyle\prod}
}}q(x_{\sigma}^{\text{new}}|x_{\sigma}^{\text{old}}).
\end{equation}
To realize MCMC with strings and branes, we keep the same
target $\pi(x_{1},...,x_{M})$, but we change the proposal kernel by
interpreting the index $\sigma$ on $x_{\sigma}$ as specifying the location of a
statistical agent in a network. Depending on the connectivity of this network,
an agent may interact with several neighboring agents $\text{Nb}(x_{\sigma})$,
so we introduce a proposal kernel:\footnote{From this perspective,
the suburban algorithm is a particular choice of ensemble MCMC
(see e.g. \cite{SwendsenWang, Geyer, AdaptiveDirectSamp,
ParallelTempering,  NealEnsemble,  AffineEnsemble}).}
\begin{equation} \label{StringKernel}
q_{\text{brane}}(x_{1}^{\text{new}},...,x_{M}^{\text{new}}|x_{1}%
^{\text{old}},...,x_{M}^{\text{old}})=\underset{\sigma=1}{\overset{M}{%
{\displaystyle\prod}
}}q_{\sigma}(x^{\text{new}}_{\sigma}|\text{Nb}(x_{\sigma}^{\text{old}})).
\end{equation}

In the above, the connectivity of the extended object specifies its overall
topology. For example, in the case of a string, i.e., a one-dimensional
extended object, the neighbors of $x_{i}$ are $x_{i-1}$, $x_{i}$, and
$x_{i+1}$. Fig. \ref{parstringbrane} depicts
the time evolution of parallel MH samplers compared with the
suburban sampler.

However, there are potentially many consistent ways to connect together the
inferences of statistical agents. From the perspective of physics, this
amounts to a notion of distance/proximity between nearest neighbors in a
brane. A physically well-motivated way to eliminate this arbitrary feature
is to allow the notion of proximity itself to \textit{fluctuate}, and for the brane
to split and join.

We view MCMC with extended strings and branes as a novel
class of ensemble samplers. By correlating the
inferences of nearest neighbors, we can expect there to be some impact
on performance. For example, the degree of
connectivity impacts the mixing rate for obtaining independent samples.
Another important feature is that because we are dealing with an
extended object, different statistical agents may become localized in
different high density regions. Provided the connectivity with neighbors is
sufficiently low, coupling these agents then has the potential to provide
a more accurate global characterization of a target distribution. Conversely,
connecting too many agents together may cause the entire collective to suffer
from \textquotedblleft groupthink\textquotedblright\ in the sense of
\cite{Heckman:2013kza}. In particular, we shall present evidence that the optimal
connectivity for a network of agents on a grid
arranged as a hypercubic lattice with some percolation (i.e., we allow for broken links)
occurs at a critical effective dimension:
\begin{equation}\label{crit}
d_{\mathrm{eff}} \sim 1
\end{equation}
where $2 d_{\mathrm{eff}} $ is the average number of neighbors.

To summarize: With too few friends one drifts into oblivion,
but with too many friends one becomes a boring conformist.

Turning our discussion around, one can view this paper as providing a concrete way to study
the physics of branes with a strongly fluctuating worldvolume, that is,
the non-perturbative regime of string theory.

The Appendices provide some additional details (see also \cite{toappear}).
The suburban code is available at \texttt{https://gitlab.com/suburban/suburban}.

\vspace{-3mm}

\section{MCMC with Strings and Branes\label{sec:MCEXTEND}}

\vspace{-3mm}

One of the main ideas we shall develop in this paper is MCMC methods for
extended objects. In an MCMC\ algorithm we produce a sequence of \textquotedblleft
timesteps\textquotedblright\ $x^{(1)},...,x^{(t)},...,x^{(N)}$ which can be
viewed as the motion of a point particle exploring a target space $\Omega$. More
formally, this sequence of points defines the \textquotedblleft
worldline\textquotedblright\ for a particle, and consequently a map:
\begin{equation}
t\mapsto x(t).
\end{equation}
For an extended object with $d$ spatial directions,
we get a map from a \textquotedblleft worldvolume\textquotedblright\ to the
target:
\begin{equation}
(t,\sigma_{1},...,\sigma_{d})\mapsto x(t,\sigma_{1},...,\sigma_{d}).
\end{equation}
The special cases $d=0$ and $d=1$ respectively denote a point particle and string.

The general physical intuition is that minus the log of the target distribution $\pi(x) = \exp(-V(x))$
defines a potential energy, and minus the log of the Markov chain transition probability
$T(x \rightarrow x^{\prime}) = \exp(-K(x,x^{\prime}))$ is a kinetic energy. The key point is that
statistical field theory in $d+1$ Euclidean dimensions strongly depends on the number of dimensions.
For example, the two-point function for a free Gaussian field $x(\sigma)$ with $\sigma \in \mathbb{R}^{d+1}$
is:
\begin{equation}\label{twopoint}
\mathbb{E} (x(\sigma) x(0) )
\sim 1 / \vert \vert \sigma \vert \vert^{d - 1}
\end{equation}
where in the case of $d = 1$, the two-point function is
$\log \vert \vert \sigma \vert \vert$. For $d \leq 1$,
a random field explores its surroundings at large $\sigma$,
but the overall variance decreases as
$d \rightarrow 1$. For $d > 1$, however, ``groupthink'' sets in
and the ensemble less quickly explores its surroundings. This
suggests a special role for stringlike objects \cite{Heckman:2013kza}.

In a theory of quantum gravity (such as string theory)
it is also physically natural to let the
proximity of nearest neighbors fluctuate. So, we introduce
an ensemble of random graphs $\mathcal{A}$.
For example, for a $d$-dimensional toroidal hypercubic lattice, introduce
$m$ lattice sites along a spatial direction so that $m^d = M$ is the total number of agents.
For a hypercubic lattice in $d$ dimensions, we define the
ensemble of random graphs for a brane $\mathcal{A}_{\text{brane}}(p_{\text{join}})$
as one in which we have a random shuffling of the agents, and in which a
given link in a $d$-dimensional hypercubic lattice is active with probability
$p_{\text{join}}$. We can also consider more general ensembles of adjacency matrices. For example,
the Erd\"{o}s-Renyi ensemble $\mathcal{A}_{\text{ER}}(p_{\text{join}})$
has an edge between any two nodes with probability $p_{\text{join}}$.
We also introduce the notion of an effective dimension
which depends on the average number of neighbors:
\begin{equation}
d_{\mathrm{eff}} = n_{\mathrm{avg}} / 2,
\end{equation}
which need not be an integer.

\vspace{-3mm}

\section{The Suburban Algorithm \label{sec:SUBURBAN}}

\vspace{-3mm}

We now present the suburban algorithm. For ease of
exposition, we shall present the case of a 1D target.
The generalization to a $D$-dimensional target is straightforward, and we can take MH within a Gibbs sampler,
or a sampler with joint variables in which all $D$ dimensions update simultaneously.

To avoid overloading the notation,
we shall write $\mathcal{X}^{(t)}\equiv\left\{  x_{1}^{(t)},...,x_{M}^{(t)}\right\}  $
for the current state of the grid. Instead of directly sampling from $\pi(x)$, we introduce
multiple copies of the target and sample from the joint distribution $\pi(\mathcal{X}) = \pi(x_{1}) ... \pi(x_{M})$
using MH sampling with proposal kernel $q_{\text{brane}}(x_{1}^{\text{new}},...,x_{M}^{\text{new}}|x_{1}%
^{\text{old}},...,x_{M}^{\text{old}}) = q(\mathcal{X}^{(\text{new})} | \mathcal{X}^{(\text{old})} , A)$,
where $A$ denotes the adjacency matrix. If the system is in a state $\mathcal{X}^{(t)}$, with
adjacency matrix $A^{(t)}$, we pick a new state according to the MH\ update rule with a proposal
kernel which depends on both these inputs.
The MH acceptance probability is:
\begin{equation}
a\left(\mathcal{X}^{\text{new}}| \mathcal{X}^{\text{old}} , A \right)
=\min\left(  1,\frac{q(\mathcal{X}^{\text{old}}|\mathcal{X}^{\text{new}}, A)}{q(\mathcal{X}^{\text{new}}|\mathcal{X}^{\text{old}}, A)}
\frac{\pi\left(  \mathcal{X}^{\text{new}}\right)  }{\pi\left(\mathcal{X}^{\text{old}}\right)}\right)
\end{equation}
This leads us to algorithm \ref{alg:suburban}.

\begin{algorithm}[t!]
\begin{algorithmic}
    \STATE Randomly Initialize $\mathcal{X}^{(0)}$ and $A^{(0)}$
    \STATE \textbf{for} $t = 0 \textrm{ to } N-1$ \textbf{do}
    \STATE \ \ \ \ $\mathcal{X}^{(\ast)} \gets$ sample from $q(\mathcal{X} | \mathcal{X}^{(t)}, A^{(t)})$
    \STATE \ \ \ \ accept with probability $a(\mathcal{X}^{\ast} | \mathcal{X}^{(t)}, A^{(t)})$
    \STATE \ \ \ \ \ \ \ \ \textbf{if} accept = true \textbf{then}
    \STATE \ \ \ \ \ \ \ \ \ \ \ \ $\mathcal{X}^{(t+1)} \gets \mathcal{X}^{(\ast)}$
    \STATE \ \ \ \ \ \ \ \ \textbf{else}
    \STATE \ \ \ \ \ \ \ \ \ \ \ \ $\mathcal{X}^{(t+1)} \gets \mathcal{X}^{(t)}$
    \STATE \ \ \ \ $A^{(t + 1)} \gets$ draw from $\mathcal{A}$
    \RETURN $\mathcal{X}^{(1)},...,\mathcal{X}^{(N)}$
\end{algorithmic}
\caption{Suburban Sampler}
\label{alg:suburban}
\end{algorithm}

Some of these steps can be parallelized whilst retaining detailed balance.
For example we could pick a coloring of a graph and then perform an update for all nodes of a particular color whilst holding fixed the
rest. We can also stochastically evolve the adjacency matrices.

Now, having collected a sequence of values $\mathcal{X}^{(1)},...,\mathcal{X}%
^{(N)}$, we can interpret this as $N\times M$ samples of the original
distribution $\pi(x)$. As standard for MCMC\ methods, we can then calculate
quantities of interest such as the mean:
\begin{equation}
\overline{x} \simeq \frac{1}{N M}\underset{\sigma,t}{%
{\displaystyle\sum}
} x_{\sigma}^{(t)}
\end{equation}
as well as higher order moments.

Let us discuss the reason we expect our sampler to converge to the correct
posterior distribution. First note that although we are modifying
the proposal kernel at each time step (i.e., by introducing a different
adjacency matrix $A\in\mathcal{A}$), this modification is independent of the
current state of the system. So, it cannot impact the eventual posterior
distribution we obtain. Second, we observe that since we are just performing a
specific kind of MH\ sampling routine for the distribution $\pi(x_{1}%
,...,x_{M})$, we expect to converge to the correct posterior distribution.
But, since the variables $x_{1},...,x_{M}$ are all independent, this
is tantamount to having also sampled multiple times from $\pi(x)$. The caveat is that we need the sampler
to actually wander around during its random walk; $d \leq 1$ is typically necessary to prevent ``groupthink.''

\vspace{-3mm}

\subsection{Implementation}

\vspace{-3mm}

To accommodate a flexible framework for
prototyping, we have implemented the suburban algorithm in the probabilistic
programming language \texttt{Dimple} \cite{Dimple}.

For practical purposes we take a fairly large burn-in cut, discarding the first $10\%$ of samples from a run.
We always perform Gibbs sampling over the $M$ agents. For
MH within Gibbs sampling over a $D$-dimensional target, we thus get a Gibbs schedule with $D \times M$ updates for each time
step. For a joint sampler, the Gibbs schedule consists of just $M$ updates.

The specific choice of $q_{\sigma}(x_{\sigma}|\text{Nb}(x_{\sigma}))$
for eqn. (\ref{StringKernel}) is motivated by having a free Gaussian field on a fluctuating graph topology:
\begin{align}\label{ColonelKlink}
\propto \exp\left(  -\alpha_{\sigma}\left(
D_{t}x_{\sigma}\right)  ^{2}-\underset{n(\sigma)}{\sum}\beta\left(  D_{t}x_{\sigma
} - D_{n(\sigma)}x_{\sigma}\right)  ^{2}\right)
\end{align}
with:
\begin{align}
D_{t}x_{\sigma} &= x_{\sigma}^{(t+1)} - x_{\sigma}^{(t)}\\
D_{n(\sigma)}x_{\sigma} &= x_{n(\sigma)}^{(t)} - x_{\sigma}^{(t)}
\end{align}
for $n(\sigma)$ a neighbor of $\sigma$ on the graph defined by the adjacency matrix. Additionally,
we set the hyperparameters for the kernel as:
\begin{equation}
\alpha_{\sigma}=2\beta-n_{\sigma}^{\mathrm{tot}}\beta,
\end{equation}
that is, we take an adaptive value for $\alpha_{\sigma}$ specified by the
number of nearest neighbors joined to $x_{\sigma}$. This condition leads to a well-behaved
continuum limit on a fully connected hypercubic lattice.

\vspace{-3mm}

\section{Numerical Experiments \label{sec:EXPERIMENT}}

\vspace{-3mm}

In most cases, we consider MH within Gibbs sampling, though we also
consider the case where joint variables are sampled, that is, pure MH.
Rather than perform error analysis within a single long MCMC run, we opt to take multiple independent trials of
each MCMC run in which we vary the hyperparameters of the sampler such as the
overall topology and average degree of connectivity of the sampler. Though this leads to
less efficient statistical estimators, it has the virtue of allowing us to easily
compare the performance of different algorithms, i.e., as we vary the continuous and discrete
hyperparameters of the suburban algorithm. We take $M = 81 = 9^2 = 3^4$ to compare
different grid topologies. We have also compared performance with parallel slice (within Gibbs)
samplers \cite{SliceSampler} to ensure that our performance is comparable to other benchmarks.

To gauge accuracy, we collect the inferred mean and covariance matrix.  We then compute
the distance to the true values:
\begin{align}
d_{\text{mean}}  &  \equiv\left\Vert \mu_{\text{inf}%
}- \mu_{\text{true}}\right\Vert \\
d_{\text{cov}}  &  \equiv\left(  \text{Tr}\left(  \left(  \Sigma_{\text{inf}%
}-\Sigma_{\text{true}}\right)  \cdot\left(  \Sigma_{\text{inf}}-\Sigma
_{\text{true}}\right)  ^{T}\right)  \right)  ^{1/2}.
\end{align}

We also collect performance metrics from the
MCMC\ runs such as the rejection rate.
A typical rule of thumb is that for targets with no large free energy barriers, a rejection
rate of somewhere between $50\%-80\%$ is acceptable
(see e.g., \cite{Gelman:1997}). We also collect the integrated auto-correlation time for
the \textquotedblleft energy\textquotedblright\ of the distribution:
\begin{equation}
V=-\log \pi(x_{1},...,x_{M}),
\end{equation}
by collecting the values $V^{(1)},...,V^{(N)}$. For $-N<k<N$, we evaluate:
\begin{equation}
c(k)\equiv\left\{
\begin{array}
[c]{c}%
\frac{1}{N}\underset{t=1}{\overset{N-k}{%
{\displaystyle\sum}
}}\left(  V^{(t)}-\overline{V}\right)  \left(  V^{(t+k)}-\overline{V}\right)  \text{
\ \ \ \ }k\geq0\\
\frac{1}{N}\underset{t=1}{\overset{N+k}{%
{\displaystyle\sum}
}}\left(  V^{(t)}-\overline{V}\right)  \left(  V^{(t-k)}-\overline{V}\right)  \text{
\ \ \ \ }k<0
\end{array}
\right\}  ,
\end{equation}
and then extract the integrated auto-correlation time:
\begin{equation}
\tau_{\text{dec}}\equiv\underset{-N<k<N}{\sum}\left(  1-\left\vert \frac{k}%
{N}\right\vert \right)  \left\vert \frac{c(k)}{c(0)}\right\vert
\end{equation}
we also refer to this as the \textquotedblleft decay time\textquotedblright
as it reflects how quickly the chain mixes. For this observable we
include all samples (no burn-in).

To extract numerical estimates we perform $T$ independent trials with random
initialization for each agent on $[-100, + 100 ]^{D}$. We present all plots with a $3$-sigma level standard
error around the mean value from these trials. In practice, we
typically find acceptable error bars for $T = 100$ and $T = 1000$ trials.

\vspace{-3mm}

\subsection{Effective Connectivity \label{sec:DIMENSION}}

\vspace{-3mm}

Perhaps the single most important feature of the suburban algorithm is that it
correlates the inferences drawn by nearest neighbors on a grid. Quite strikingly, we find
that the effective dimension rather than the overall topology of the grid
plays the dominant role in the performance of the algorithm.

We illustrate this point with a class of target distribution examples which we refer to as
\textquotedblleft symmetric mixtures.\textquotedblright For a fixed choice of
$D$ the number of target space dimensions, we introduce a mixture model
consisting of $2D$ equal weight components, each of which is a normal distribution
with means and covariance matrices:
\begin{equation}
\mu_{i}^{(\pm,j)} = \pm \mu\times\delta_{i}^{j}\,\,\,\,\,\,\Sigma^{(\pm, i)} =\sigma^{2} \times \mathbb{I}_{D \times D},
\end{equation}
where $i,j =1,...,D$, $\delta_{i}^{j}$ is a Kronecker delta and
$\mathbb{I}_{D \times D}$ is the $D \times D$ identity matrix.
We find qualitatively similar behavior for $D=2$ and $D=10$, so we give the plots
for the $D=2$ runs with $\mu=1.5$ and $\sigma^{2}=0.25$. In this case, we have four equally
weighted components of our mixture model, and there is a free energy barrier
separating these centers.

As a first class of tests, we consider sampling with different topology grids for
$N=10,000$ timesteps, with each grid consisting of $M=81$ agents.
For each choice of hyperparameter, we perform $T = 100$ trials.

We have scanned the value of $\beta$ in steps of factors of $10$, and
find that performance is better around $\beta=0.01$, so we focus
on this case. In all cases, we find that the values of the
observables $d_{\text{mean}}$ and $d_{\text{cov}}$ are comparable and small, indicating
reasonable convergence.

\begin{figure}[ptb]%
\centering
\includegraphics[
scale = 0.50, trim = 18mm 70mm 0mm 70mm
]%
{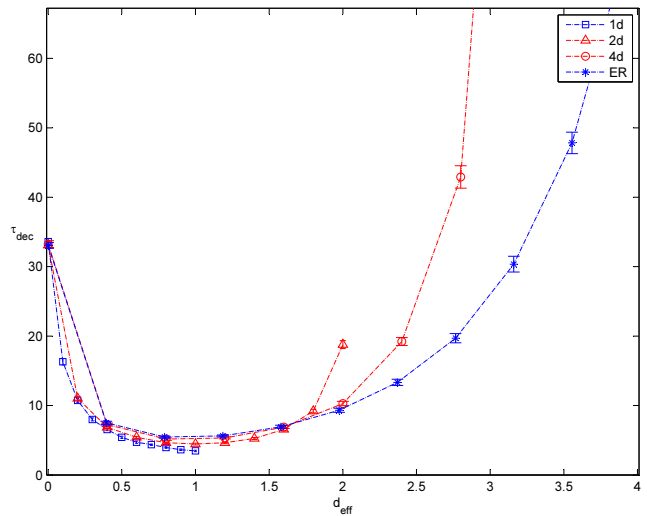}%
\caption{Plot of the mixing rate $\tau_{\text{dec}}$ as a
function of the effective dimension of the grid. All data comes from sampling
the $D = 2$ symmetric mixture model.}%
\label{twodcubetaus}%
\end{figure}

There is, however, a marked difference in the mixing rate as we vary the split / join probability
for the ensemble. In fig. \ref{twodcubetaus} we
display the values of $\tau_{\text{dec}}$ as a function of the effective
dimension dictated by the split / join rate for a given grid topology.
Quite striking is the universal behavior of
the samplers as a function of the effective dimension near $d_{\text{eff}}%
\sim1$, i.e., for connectivity similar to that of a string.
Near $d_{\text{eff}}=0$, i.e., for parallel
MH\ samplers, we also see much slower mixing rates. Once we go beyond
$d_{\text{eff}} \gtrsim 1$, the overall performance of the sampler
suffers.%

\begin{figure}[ptb]%
\centering
\includegraphics[
scale = 0.50, trim = 18mm 70mm 0mm 70mm
]%
{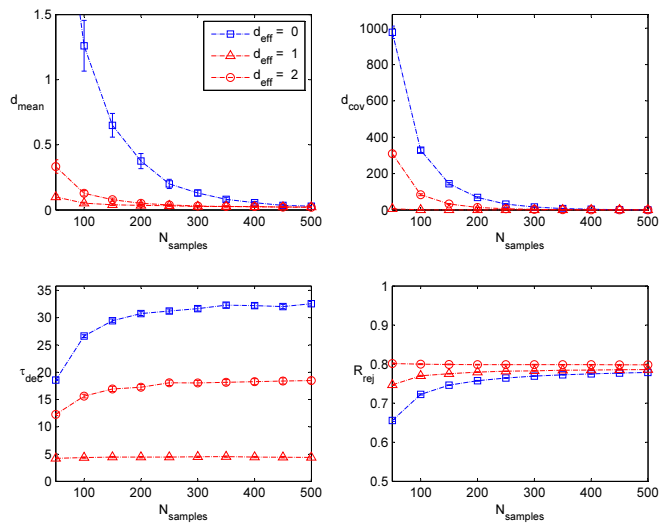}%
\caption{Plots of the 2D symmetric mixture model tests.}
\label{timetwodcubemets}%
\end{figure}

Because of this universal behavior, we shall primarily focus on
\textquotedblleft representative behavior\textquotedblright\ as
obtained from a $2d$ grid topology. In fig. \ref{timetwodcubemets} we show various performance metrics as a
function of the total number of samples. By inspection, stringlike samplers
tend to fare the best.

\vspace{-3mm}

\subsection{Random Landscapes \label{sec:LANDSCAPE}}

\vspace{-3mm}

As another example, we consider mixture models in which there is a
landscape of local maxima and minima. A priori, a compromise will need
to be struck between \textquotedblleft wandering freely\textquotedblright\
and moving more slowly around individual components of the mixture model.

\begin{figure}[t!]%
\centering
\includegraphics[
scale = 0.50, trim = 18mm 70mm 0mm 70mm
]%
{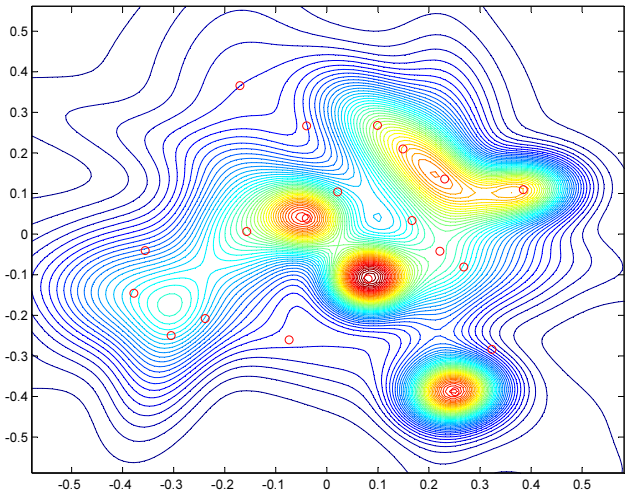}%
\caption{Contour plot of the random mixture model with twenty components
and random seed of $40$. Red circles denote centers of individual components.}%
\label{rand40plot}%
\end{figure}

We use a variant on the same random mixture model considered
in \cite{Herding}, focussing on the case of $20$ Gaussian
mixtures with relative weights randomly drawn from the uniform
distribution on $[0,1]$. Compared with \cite{Herding},
we take the parameters \texttt{stdmu} $=0.4$, \texttt{stdsig} $=10.0$. We do this primarily to achieve convergence
for the samplers in a reasonable amount of time. The different mixture models
are obtained by setting the random seed in the code of \cite{Herding} to different
values (see fig. \ref{rand40plot}).

By design, we have chosen our domain for the random variables
so that the brane tension $\beta = 0.01$ should give a roughly comparable
class of length scales for the target distribution. Since the overall topology
of the grid does not appear to affect the qualitative behavior of the sampler,
we have also focussed on the case of a $2d$ grid topology with shuffling and percolation.
For each choice of hyperparameter, we perform $T = 100$ trials.

\begin{figure}[t!]%
\centering
\includegraphics[
scale = 0.50, trim = 18mm 70mm 0mm 70mm
]%
{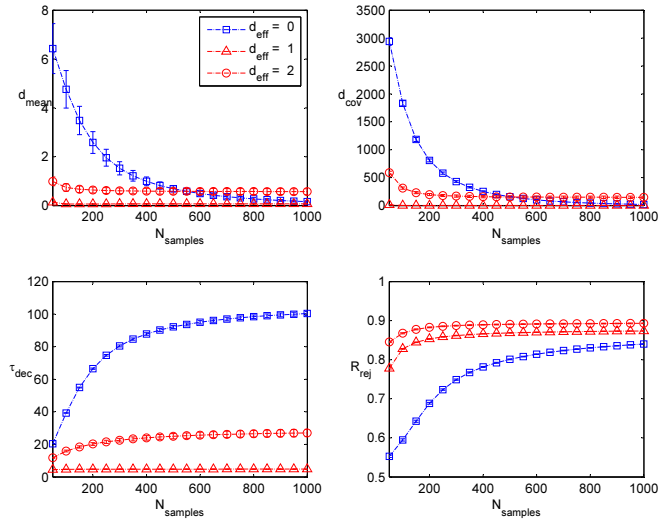}%
\caption{Plots of the random seed $40$ landscape tests.}%
\label{rand40b0p01mets}%
\end{figure}

The random seed $40$ model gives representative behavior.
The stringlike sampler is faster and more accurate than the parallel MH sampler,
while a $d_{\text{eff}} \sim 2$ sampler suffers from ``groupthink,''
settling in an incorrect metastable configuration.

\vspace{-3mm}

\subsection{Banana Distribution \label{sec:BANANA}}

\vspace{-3mm}

It is also of interest to consider distributions concentrated on a
lower-dimensional subspace such as the two-dimensional \textquotedblleft banana
distribution\textquotedblright:
\begin{equation}
\pi_{\text{banana}}(x,y)=\frac{1}{10\pi}\exp\left(  -(x-1)^{2}-100(y-x^{2}%
)^{2}\right)  .
\end{equation}
This distribution is often used as a performance test of various optimization
algorithms.

We focus on a suburban sampler with joint variables, taking different
grid topologies for the statistical agents and then perform a
sweep over different values of the hyperparemeters $\beta$ and $d_{\text{eff}%
}$, performing $T = 100$ trials for each case.

\begin{figure}[t!]%
\centering
\includegraphics[
scale = 0.50, trim = 18mm 70mm 0mm 70mm
]%
{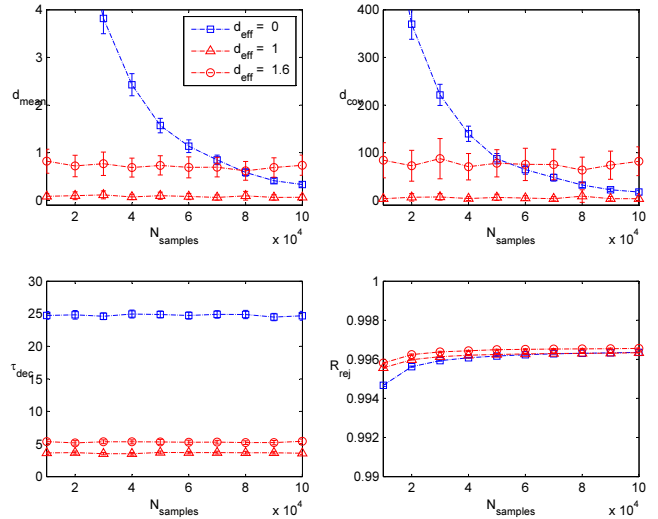}%
\caption{Plots of the banana distribution tests.}%
\label{time2dbananamets}%
\end{figure}

We present the representative case of a $2d$ grid, and further specialize
to the tuned case of $\beta=0.01$. Fig. \ref{time2dbananamets}
shows that parallel samplers ($d_{eff} = 0$) and collectives with
groupthink $(d_{eff} > 1)$ both fare worse than a stringlike sampler.

\vspace{-3mm}

\subsection{Free Energy Barriers \label{sec:FREEBALL}}

\vspace{-3mm}

The extended nature of the suburban sampler also suggests that for target distributions
with various disconnected \textquotedblleft deep pockets,\textquotedblright%
\ different pieces of the ensemble can wander over to different regions. We consider
a mixture model with two Gaussian components:
\begin{equation}
\pi_{\text{GMM}}(x)=\frac{3}{4}\mathcal{N}(x|\mu^{(+)},\Sigma)+\frac{1}%
{4}\mathcal{N}(x|\mu^{(-)},\Sigma),
\end{equation}
with:
\begin{equation}
\mu^{(\pm)} =(\pm L_{\text{barrier}},0,...,0) \,\,\,\,\,\, \Sigma =\sigma^{2}\times \mathbb{I}_{D\times D},
\end{equation}
where we vary $L_{\text{barrier}}$ and hold fixed $\sigma = 0.25$. We take
$N_{\text{samples}}=1000$ samples with $M=81$ agents on a $2d$ grid with $\beta = 0.01$, performing
$T = 1000$ independent trials. Since we use MH within Gibbs, we do not find
much decrease in performance in comparing the $D=2$ and $D=10$ free energy barrier
tests.

\begin{figure}[b!]%
\centering
\includegraphics[
scale = 0.50, trim = 18mm 65mm 0mm 65mm
]%
{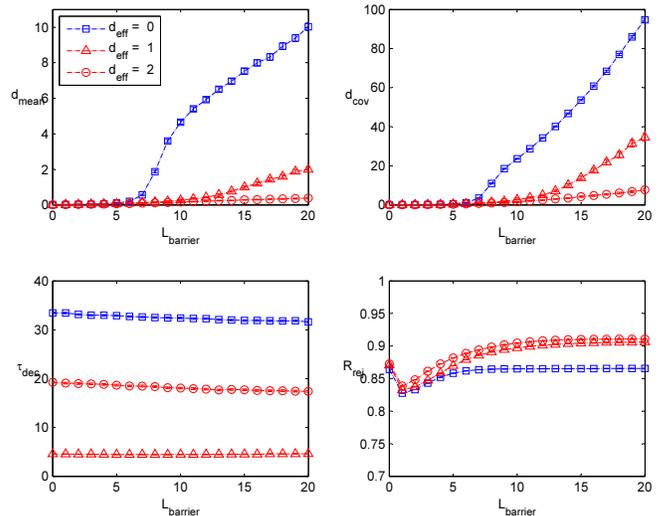}%
\caption{Plots of the 2D free energy barrier tests.}%
\label{2dfreebarrierplotmet}%
\end{figure}

Fig. \ref{2dfreebarrierplotmet} shows that parallel samplers fare worse than
the extended objects. The $d_{\text{eff}}=2$ runs are sometimes more accurate,
but mix slower than for $d_{\text{eff}}=1$. After thinning samples, the former
runs will be less accurate.

\vspace{-3mm}

\section*{Acknowledgements}

\vspace{-3mm}

JJH thanks D. Krohn for collaboration at an early stage. We thank
J.A. Barandes, C. Barber, C. Freer, M. Freytsis, J.J. Heckman Sr., A. Murugan, P. Oreto, R. Yang, and J. Yedidia for helpful discussions.
The work of JJH is supported by NSF CAREER grant PHY-1452037.
JJH also acknowledges support from the Bahnson Fund as well as the
R.~J. Reynolds Industries, Inc. Junior Faculty Development Award at UNC Chapel Hill.


\vspace{-3mm}

\appendix

\vspace{-3mm}

\section{Path Integrals for MCMC \label{sec:MCEXTEND}}

\vspace{-3mm}

In this Appendix we give a path integral formulation for MCMC with extended objects.
For additional background on path integrals in statistical field theory, see \cite{Peskin:1995ev, Wipf:2013vp}.
In what follows, we denote the random variable as $X$ with outcome $x$ on a target space $\Omega$
with measure $dx$. We consider sampling from a probability density $\pi(x)$. In accord with physical
intuition, we view $- \log \pi(x)= V(x)$ as a potential energy.

In general, our aim is to discover the structure of $\pi(x)$ by using some
sampling algorithm to produce a sequence of values $x^{(1)},...,x^{(N)}$. A
quantity of interest is the expected value of $\pi(x)$ with respect to a given
probability distribution of paths. This helps in telling us the relative speed of convergence and the
mixing rate. To study this, it is helpful to evaluate the expectation value of
the quantity:%
\begin{equation} \label{productpot}
\underset{i=1}{\overset{N}{%
{\displaystyle\prod}
}}\exp(- V(x^{(i)}))
\end{equation}
with respect to a given path generated by our sampler.

In more general terms, the reason to be interested in this expectation value comes
from the statistical mechanical interpretation of statistical
inference \cite{Balasubramanian:1996bn, Heckman:2013kza}: There is a
competition between staying in high likelihood regions (minimizing the potential), and exploring more of the distribution (maximizing
entropy). The tradeoff between the two is neatly captured by the path integral formalism: It tells us about
a particle moving in a potential $V(x)$, and subject to a thermal background, as specified by the
choice of probability measure over possible paths. Indeed, we will view this probability measure as defining a ``kinetic energy''
in the sense that at each time step, we apply a random kick to the trajectory of the particle,
as dictated by its contact with a thermal reservoir.

Along these lines, if we have an MCMC\ sampler with transition
probabilities $T(x^{(i)}\rightarrow x^{(i+1)})$, marginalizing
over the intermediate values yields the expected value
of line (\ref{productpot}):
\begin{equation}
\mathcal{Z} = \int [dx^{(t)}] \left(
\underset{i=0}{\overset{N-1}{%
{\displaystyle\prod}
}}T(x^{(i)}\rightarrow x^{(i+1)})e^{-V(x^{(i+1)})}\right)
\end{equation}
where we have introduced the measure factor $[dx^{(t)}] = dx^{(1)} ... dx^{(N)}$.
We would like to interpret $V(x)$ as the potential energy and $-\log
T(x^{(i)}\rightarrow x^{(i+1)})$ as a kinetic energy:
\begin{equation}
V = -\log\pi \text{ \ \ and \ \ }K = -\log T,
\end{equation}

We now observe that our expectation value has the form of a well-known object
in physics:
\begin{equation}
\mathcal{Z}(x^{\text{begin}}\rightarrow x^{\text{end}})=\underset{\text{begin}%
}{\overset{\text{end}}{%
{\displaystyle\int}
}}[dx^{(t)}] e^{- \underset{t}{\sum} L^{(E)}[x^{(t)}]},
\end{equation}
A path integral! Here the Euclidean signature Lagrangian is:
\begin{equation}
L^{(E)}[x^{(t)}]=K+V.\text{ }%
\end{equation}
Since we shall also be taking the number of timesteps to be very large, we
make the Riemann sum approximation and introduce the rescaled Lagrangian
density:%
\begin{equation}
\frac{1}{N}\underset{t}{\sum}\mapsto\int dt,\text{ \ \ }NL^{(E)}%
\mapsto\mathcal{L}^{(E)}%
\end{equation}
so that we can write our process as:
\begin{equation}
\mathcal{Z}(x^{\text{begin}}\rightarrow x^{\text{end}})=\int[dx] e^{-\int dt\mathcal{L}^{(E)}[x(t)]}  ,
\end{equation}
where by abuse of notation, we use the same variable $t$ to reference both the
discretized timestep as well as its continuum counterpart.

To give further justification for this terminology, consider now the specific
case of the Metropolis-Hastings algorithm. In this case, we have a proposal
kernel $q(x^{\prime}|x)$, and acceptance probability:%
\begin{equation}
a(x^{\prime}|x)=\min\left(  1,\frac{q(x|x^{\prime})}{q(x^{\prime}|x)}\frac
{\pi(x^{\prime})}{\pi(x)}\right)  .
\end{equation}
The total transmission probability is then given by a sum of two terms. One is
given by $a(x^{\prime}|x)q(x^{\prime}|x)$, i.e., we accept the new sample. We
also sometimes reject the sample, i.e., we keep the same value as before:%
\begin{equation}
T(x\rightarrow x^{\prime})=r\times\delta(x-x^{\prime})+a(x^{\prime
}|x)q(x^{\prime}|x),
\end{equation}
where $\delta(x - x^{\prime})$ is the Dirac delta function, and we have introduced
an averaged rejection rate:
\begin{equation}
r \equiv1-\int dx^{\prime}\text{ }a(x^{\prime}|x)q(x^{\prime}|x).
\end{equation}

To gain further insight, we now approximate the mixture model $T(x\rightarrow
x^{\prime})$ by a normal distribution $q_{\text{eff}}\left(  x^{(t+1)}%
|x^{(t)}\right)  $ such that  $-\log q_{\text{eff}}\left(  x^{(t+1)}%
|x^{(t)}\right)  \sim\alpha_{\text{eff}}\left(  x^{(t+1)}-x^{(t)}\right)
^{2}$. Hence,\footnote{For example,
for a Gaussian proposal kernel with $- \log q(x^{(t+1)}|x^{(t)}) \sim \alpha (x^{(t+1)} - x^{(t)})^2$,
matching the first and second moments to $q_{\text{eff}}$ requires $\alpha_{\text{eff}} = \alpha / \overline{a}$,
with $\overline{a}$ the average acceptance rate.}
\begin{equation}\label{generalLag}
L^{(E)}[x^{(t)}]\simeq\alpha_{\text{eff}}\left(  x^{(t+1)}-x^{(t)}\right)
^{2}+V(x^{(t)})+...,
\end{equation}
where here, the \textquotedblleft...\textquotedblright\ denotes additional
correction terms which are typically suppressed by powers of $1/N$.

Our plan will be to assume a kinetic term with quadratic time derivatives, but a general
potential. The overall strength of the kinetic term will depend
on details such as the average acceptance rate. As the acceptance
rate decreases, $\alpha_{\text{eff}}$ increases and the sampled
values all concentrate together.

We now turn to the generalization of the above concepts for strings and
branes, i.e., extended objects. Introduce $M$ copies of the original distribution, and
consider the related joint distribution:%
\begin{equation}
\pi(x_{1},...,x_{M})=\pi(x_{1})...\pi(x_{M})\text{.}%
\end{equation}
If we keep the proposal kernel unchanged, we can simply describe the evolution
of $M$ independent point particles exploring an enlarged target space:%
\begin{equation}
\Omega_{\text{enlarged}}= \Omega^{M}=\underset{M}{\underbrace{\Omega \times...\times \Omega}
}\text{.}%
\end{equation}
If we also view the individual statistical agents on the worldvolume as
indistinguishable, we can also consider quotienting by the symmetric group on
$M$ letters, $S_{M}$:%
\begin{equation}
\Omega^{\mathcal{S}}_{\text{enlarged}}=X^{M}/S_{M}\text{.}%
\end{equation}

Of course, we are also free to consider a more general proposal kernel in
which we correlate these values. Viewed in this way, an extended object is a
single point particle, but on an enlarged target space. The precise way in
which we correlate entries across a grid will in turn dictate the type of
extended object.

Indeed, much of the path integral formalism carries over unchanged. The only
difference is that now, we must also keep track of the spatial extent of our
object. So, we again introduce a potential energy $V$ and a kinetic energy $K$:
\begin{equation}
V = -\log\pi \text{ \ \ and \ \ }K = -\log T,
\end{equation}
and a Euclidean signature Lagrangian density:
\begin{equation}
L^{(E)}[x(t,\sigma_{A})]=K+V,
\end{equation}
where here, $\sigma_{A}$ indexes locations on the extended object, and the
subscript $A$ makes implicit reference to the adjacency on the graph. In a similar notation,
the expected value is now:
\begin{equation}
\mathcal{Z}(x_{\text{begin}}\rightarrow x_{\text{end}} | A)=%
{\displaystyle\int}
[dx]\text{ } e^{-\underset{t}{\sum}\underset{\sigma}{\sum
}L^{(E)}[x(t,\sigma_{A})]}.
\end{equation}
Since we shall also be taking the number of time steps and agents to be large, we
again make the Riemann sum approximation:
\begin{equation}
\frac{1}{N}\underset{t}{\sum}\mapsto\int dt,\text{ \ \ }\frac{1}%
{M}\underset{\sigma}{\sum}\mapsto\int d\sigma_{A}\text{ \ \ }NML^{(E)}%
\mapsto\mathcal{L}^{(E)}
\end{equation}
so that:
\begin{equation}
\mathcal{Z}(x^{\text{begin}}\rightarrow x^{\text{end}} | A)=\int%
[dx] e^{ -\int dtd\sigma_{A}\text{ }\mathcal{L}%
^{(E)}[x(t,\sigma_{A})]} ,
\end{equation}
in the obvious notation.

So far, we have held fixed a particular adjacency matrix. This is somewhat arbitrary, and physical considerations suggest
a natural generalization where we sum over a statistical ensemble of choices.
One can loosely refer to this splitting and joining of connectivity
as \textquotedblleft incorporating gravity\textquotedblright\ into the dynamics of the extended object, because
it can change the notion of which statistical agents are nearest
neighbors.\footnote{It is not quite gravity in the worldvolume theory, because
there is a priori no guarantee that our sum over different graph topologies
will have a smooth semi-classical limit. Nevertheless, summing over different
ways to connect the statistical agents conveys the main point that the
proximity of any two agents can change.} Along these lines, we incorporate an
ensemble $\mathcal{A}$ of possible adjacency matrices, with some prescribed
probability to draw a given adjacency matrix. The topology
of an extended object dictates a choice of statistical ensemble $\mathcal{A}$.

Since we evolve forward in
discretized time steps, we can in principle have a sequence of such matrices
$A^{(1)},...,A^{(N)}$, one for each timestep. For each draw of an adjacency
matrix, the notion of nearest neighbor will change, which we denote by writing
$\sigma_{A(t)}$, that is, we make implicit reference to the connectivity of
nearest neighbors. Marginalizing over the choice of adjacency matrix, we get:%
\begin{equation}
\mathcal{Z}(x_{\text{begin}}\rightarrow x_{\text{end}})=%
{\displaystyle\int}
[dx][dA]\text{ } e^{-\underset{t}{\sum}\underset{\sigma
}{\sum}L^{(E)}[x(t,\sigma_{A(t)})]},
\end{equation}
where now the integral involves summing over multiple ensembles:
the spatial and temporal values with measure factor $dx_{\sigma}^{(t)}$, as well as the choice
of a random matrix from the ensemble $dA^{(t)}$ (one such integral for each
timestep). At a very general level, one can view the adjacency matrix as
adding additional auxiliary random variables to the process. So in this sense,
it is simply part of the definition of the proposal kernel.

\vspace{-3mm}

\subsection{Dimensions and Correlations} \label{ssec:SCALING}

\vspace{-3mm}

Following some of the general considerations outlined in reference \cite{Heckman:2013kza},
we now discuss the extent to which the extended nature of such objects
plays a role in statistical inference and in particular MCMC.

To keep our discussion from becoming overly general, we
specialize to the case of a hypercubic lattice of agents in $d$ spatial dimensions
arranged on a torus, and we denote a location on the grid by a $d$-component
vector $\sigma$. We can allow for the possibility of a fluctuating worldvolume
by making the crude substitution $d \mapsto d_{\text{eff}}$.

Consider the Gaussian proposal kernel of line (\ref{ColonelKlink}).
In a large lattice, we approximate the finite differences in one
of the $d$ spatial directions by derivatives of continuous functions.
Expanding in this limit, various cross-terms cancel
and we get for the proposal kernel:
\begin{equation}
\propto\exp\left(  -2\beta
\underset{\sigma}{%
{\displaystyle\sum}
}\left(  \left(  D_{t}x_{\sigma}\right)  ^{2}%
+\underset{k=1}{\overset{d}{\sum}}\left(  D_{k}x_{\sigma
}\right)  ^{2}\right)  \right)  .
\end{equation}
where $D_{k}$ denotes a finite difference in
the $k^{th}$ spatial component of the $d$-dimensional lattice.

Just as in the case of the point particle, the transition rate defines
a kinetic energy quadratic in derivatives (to leading order),
with an effective strength dictated by the overall acceptance rate.

One of the things we would most like to understand is the extent to which an
extended object can explore the hills and valleys
of $V$. We perform a perturbative analysis, at first viewing $V$ as a small correction to the
Lagrangian. Starting from some fixed position
$x_{\ast}$, consider the expansion of $V$ around this point:%
\begin{equation}
V(x)=V(x_{\ast})+ V^{\prime}(x_{\ast}) (x - x_{\ast}) + \frac{V^{\prime\prime}(x_{\ast})}{2} (x - x_{\ast})^2 + ...,
\label{PotentialExpand}%
\end{equation}
Each of the derivatives of $V(x)$ reveals another characteristic feature length of
$V(x)$. These feature lengths are specified by the values of the moments
for the distribution $\pi(x)$.

When $V=0$, there is a well-known behavior of correlation functions which is given by eqn. (\ref{twopoint}).\footnote{One
way to obtain this scaling relation is to observe that the Fourier
transform of $1/k^2$ in $d+1$ dimensions exhibits the requisite power law behavior.}
There is thus a rather sharp change in the inferential powers of an extended object above and below
$d_{\text{eff}} \sim 1$.

To understand the impact of a non-trivial potential,
we introduce the notion of a ``scaling dimension'' for $x(t,\sigma)$ and its derivatives.
This is a well-known notion, see \cite{DiFrancesco:1997nk} for a review.
Just as we assign a notion of proximity in space and time to agents on a grid,
we can also ask how rescaling all distances on the grid via:
\begin{equation}\label{rescaler}
N \mapsto \lambda N \,\,\, M \mapsto \lambda^{d} M
\end{equation}
impacts the structure of our continuum theory Lagrangian.
The key point is that provided $N$ and $M$ have been taken sufficiently
large, or alternatively we take $\lambda$ sufficiently large, we
do not expect there to be any impact on the physical interpretation.

Unpacking this statement naturally leads us to the notion of a
scaling dimension for $x(t, \sigma)$ itself. Observe that
rescaling the number of samples and number of agents in
line (\ref{rescaler}) can be interpreted equivalently as holding
fixed $N$ and $M$, but rescaling $t$ and $\sigma$:
\begin{equation}
(t,\sigma) \mapsto (\lambda t , \lambda \sigma).
\end{equation}
Now, for our kinetic term to remain invariant, we need to \textit{also} rescale $x(t,\sigma)$:
\begin{equation}
x(t, \sigma) \mapsto \lambda^{- \Delta} x( \lambda t , \lambda \sigma).
\end{equation}
The exponent $\Delta$ is often referred to as the ``scaling dimension'' for $x$ obtained from ``naive dimensional analysis'' or NDA.
It is ``naive'' in the sense that when the potential $V \neq 0$ and we have strong coupling, the notion of a scaling dimension may only emerge at sufficiently long distance scales. Note that because we
are uniformly rescaling the spatial and temporal pieces of the grid, we get the same answer for the scaling dimension if we consider spatial derivatives along the grid. This assumption can also be relaxed in more general physical systems.
To illustrate, invariance of the free field action requires:
\begin{equation}\label{freedim}
\Delta = \frac{d - 1}{2}.
\end{equation}
We can also consider the behavior of a perturbation of the form $(x)^{\mu} (D x)^{\nu}$.
Applying our NDA analysis prescription, we see that under a rescaling, the contribution such a term makes to the action is:
\begin{equation}
\int dt d^{d} \sigma \text{ } (x)^{\mu} (D x)^\nu \mapsto  \lambda^{-\mu \Delta - \nu (\Delta + 1) + d + 1} \int dt d^{d} \sigma \text{ } (D x)^2,
\end{equation}
so terms of the form $(D x)^{\nu}$ for $\nu > 2$ die off
as we take $N \rightarrow \infty$, i.e., $\lambda \rightarrow \infty$. Additionally, we
see that when $d \leq 1$, we can in principle expect more general contributions of
the form $(x)^{\mu} (D x )^{\nu}$. For additional discussion on the interpretation of such contributions,
see reference \cite{Heckman:2013kza}.

Consider next possible perturbations to the potential energy.
Each successive interaction term in the potential is of the form
$x^{n}$, with scaling dimension $n(d-1)/2$. So, for $d\leq1$,
all higher order terms can impact the long distance behavior
of the correlation functions, while for $d>1$, the most relevant term is
bounded above, and the global structure of the potential will be missed.

\begin{figure}[t!]%
\centering
\includegraphics[
scale = 0.50, trim = 18mm 70mm 0mm 70mm
]%
{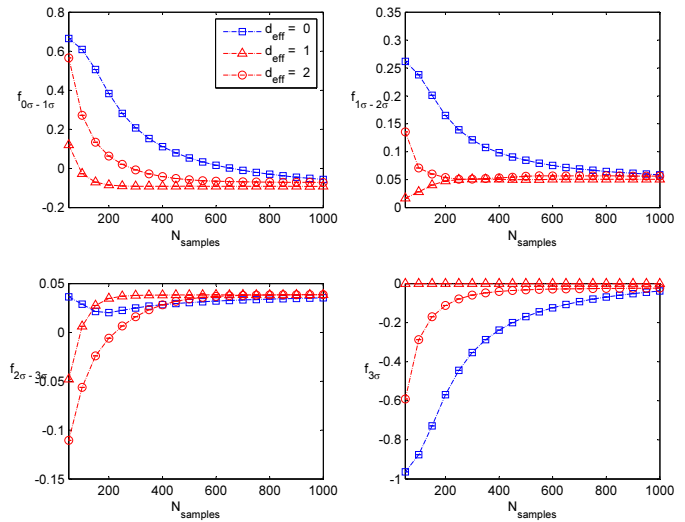}%
\caption{Plots of the convergence to the true tail statistics for suburban
samplers of the random seed $40$ landscape.}%
\label{rand40b0p01tails}%
\end{figure}

\begin{figure}[t!]%
\centering
\includegraphics[
scale = 0.50, trim = 18mm 70mm 0mm 70mm
]%
{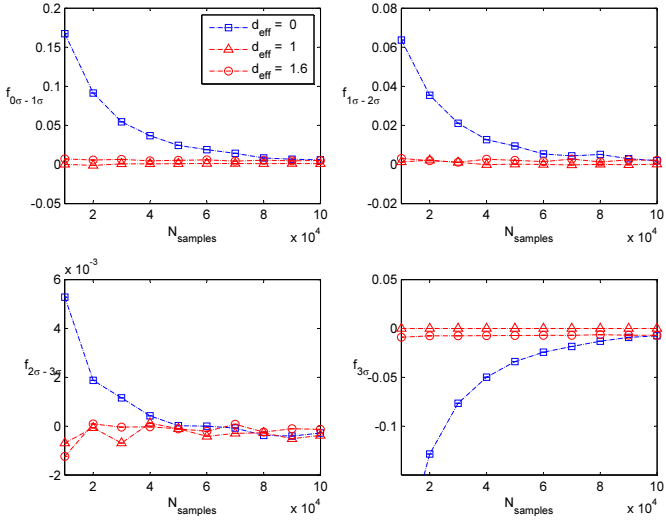}%
\caption{Plots of the convergence to the true correct tail statistics for
suburban samplers of the banana distribution.}%
\label{time2dbananatails}%
\end{figure}

\vspace{-3mm}

\section{Tail Statistics Tests}

\vspace{-3mm}

In figs. \ref{rand40b0p01tails} and \ref{time2dbananatails} we display
some tests of how well a sampler collects ``rare events,'' i.e., tail
statistics. After taking a burn-in cut with $N_{\text{total}}$ remaining samples,
we compute the number of counts in the $0\sigma-1\sigma$ region, the $1\sigma-2\sigma$
region, the $2\sigma-3\sigma$ region, and events which fall outside the
$3\sigma$ region. For each such region, we compute the difference between the
inferred and true counts and return the fraction:
\begin{equation}
f_{\text{region}}\equiv\frac{N_{\text{inf}}-N_{\text{true}}}{N_{\text{total}}%
}.
\end{equation}

\begin{figure}[t!]%
\centering
\includegraphics[
scale = 0.50, trim = 18mm 70mm 0mm 70mm
]%
{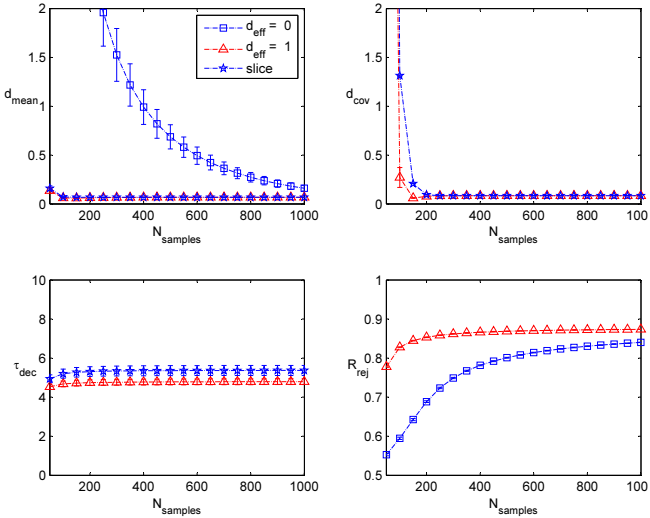}
\caption{Comparison between slice and suburban for the random seed $40$ landscape.}
\label{LandRand40SliceVsMh}%
\end{figure}

\begin{figure}[b!]%
\centering
\includegraphics[
scale = 0.50, trim = 18mm 70mm 0mm 70mm
]%
{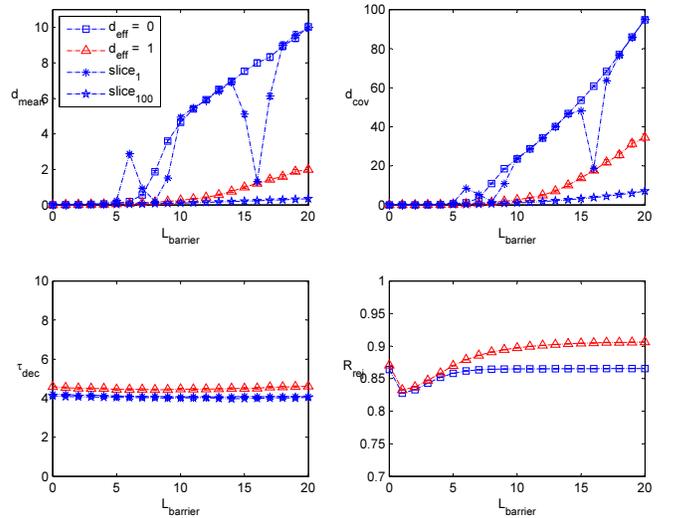}%
\caption{Comparison between slice and suburban for the free energy barrier test. We show
two different initialization widths for the slice sampler.}%
\label{2dfreebarrierplotmetNEW}%
\end{figure}


\section{Comparison with Slice} \label{app:SLICE}

\vspace{-3mm}

Figs. \ref{LandRand40SliceVsMh} and \ref{2dfreebarrierplotmetNEW}
compare the performance of a suburban sampler with
$2d$ grid topology (with $d_{\mathrm{eff}} = 1$ and $\beta = 0.01$)
with parallel slice within Gibbs sampling. We use the default implementation in \texttt{Dimple} so that
for a 1D target distribution the initial size
of the $x$-axis width is an interval of length one
containing $x_{\ast}$, and the maximum number of doublings is $10$.
A direct comparison with suburban is subtle because in slice sampling
the halting of the ``stepping out'' and ``stepping in'' loops is not fixed ahead of time.
In practice we find that for a fixed number of samples, slice typically makes several more queries to the target
distribution compared with suburban, roughly a factor of $\sim 5 - 10$. For large free energy barriers,
it is also sometimes helpful to enlarge the initialization width from $1$ to $100$ (see fig. \ref{2dfreebarrierplotmetNEW}).

\end{document}